\documentclass{article}
\usepackage{amsmath,amsthm,amssymb}
\usepackage[square,numbers]{natbib}
\usepackage[latin1]{inputenc}
\usepackage[T1]{fontenc}
\usepackage{url}
\usepackage{color}
\usepackage{epsfig}
\usepackage[normalem]{ulem}
\usepackage{listings}
\usepackage{tikz}
\usetikzlibrary{arrows,decorations.pathmorphing,backgrounds,fit,positioning,shapes.symbols,chains}
\usepackage{float}
\usepackage{hyperref}
\usepackage{wrapfig}
\usepackage{multirow}
\usepackage{algorithm2e}

\usetikzlibrary{decorations.pathmorphing}
\usetikzlibrary{arrows,shapes,trees}

\definecolor{dgreen}{rgb}{0,.6,0}

\graphicspath{{figures/}}

\title{A formal methodology for integral security design and verification of network protocols}
\author{Jesus Diaz, David Arroyo, Francisco B. Rodriguez}

\begin{document}
\maketitle

  \begin{abstract}
    We propose a methodology for verifying security properties of network protocols 
    at design level. It can be separated in two main parts: context and requirements 
    analysis and informal verification; and formal representation and procedural 
    verification. It is an iterative process where the early steps are simpler 
    than the last ones. Therefore, the effort required for detecting flaws is 
    proportional to the complexity of the associated attack. Thus, we avoid wasting 
    valuable resources for simple flaws that can be detected early in the 
    verification process. In order to illustrate the advantages provided by our 
    methodology, we also analyze three real protocols.
  \end{abstract}

\section{Introduction}
\label{sec:introduction}

With this work we would like to promote
dissemination of methodologies for the verification of 
security properties of interactive protocols. It covers the whole
protocol design process, from the specification of the requirements and the
protocol environment, to the protocol definition and its validation. Such a 
methodology will help avoiding errors due to wrong or no longer valid assumptions 
on the context in which the protocol is executed, while also preventing design 
flaws that can lead to the non-fulfillment of the security requirements. Indeed, 
there already exist plenty of tools for the formal analysis of those 
security requirements \cite{blanchet10,cryptyc}. 
However, even though there 
exist automatic tools for formal security verification, dedicating some time previous to this formal
analysis to the task of defining a security context may help to detect some initial
(but important) flaws. This way, we also prevent an unnecessary waste of time and improve the
understanding of the protocol. Like we will see, an active application of a 
methodology covering the whole design process will help reducing the impact of 
illegitimate actions over the resulting network protocols.

The  paper is structured as follows. We start in Sec. \ref{sec:related} with a brief introduction to 
existing frameworks and procedures for verification of security properties, along
with a discussion on the advantages of applying the verification procedures at
the different stages of the software life cycle. In Sec. \ref{sec:methodology} we introduce 
our methodology. In Sec. \ref{sec:cases} we apply the different phases of our 
methodology to three different protocols (MANA III, WEP-SKA and CHAT-SRP), and 
we conclude this work in Sec. \ref{sec:conclusion}, with a global perspective of 
the benefits provided by our methodology.

\section{Related work and discussion}
\label{sec:related}

Security flaws in information systems are receiving increasing attention,
as the problems and losses they can cause grow in severity \cite{sw09,bi03,yc07}. As a result,
organizations are paying more attention to the design of secure systems.
In that matter, several frameworks have been developed during the last years, 
intended to 
help in the task of creating secure software by delimiting where can security flaws 
be originated \cite{sdl,stride,jurjens04}. There are also proposals and standards defining 
different security levels that may be required depending on the context and managed 
information, and even the methods and tools used for verification \cite{cc3,mmob10}. 
While \cite{sdl,stride,cc3} are aimed for every software system and all the components 
within it, the proposal in \cite{mmob10} is centered in evaluating security in 
cryptographic protocols. Certainly, cryptographic protocols are quite important for 
the evaluation of security properties,
since they are applied when something is required to be protected. Moreover,
protocols can be seen as components that need special consideration when
evaluating security because they are subject to specific risks that can hardly
take place in other components (e.g., replay, spoofing or Man In The Middle 
attacks). In the mentioned work, several levels of security are defined, depending 
on the level of formalization applied in the different elements of the methodology 
(protocol specification, adversarial model, security properties and self-assessment
evidence), and the type of tools used in the process. Indeed, we shall refer to
them \cite[Sec. 2.1]{mmob10} for a good review of the different type of formal methods and
tools that can be used for cryptographic protocols analysis.

Besides the classification of verification tools in terms of the theory they are
based upon, there is yet another further division: the point in the software development
life cycle where they are applied. Precisely, they can be applied to check the 
design obtained after the design phase; or to check the code created during the
implementation phase. Examples of tools for verification at the design phase are
ProVerif, CryptoVerif, CertiCrypt, Cryptyc, AVISPA and Isabelle \cite{blanchet10,cryptoverif,bgz09,cryptyc,avispa,isabelle}. 
On the other hand, examples of code verifiers are \cite{bfg10,cd09,gp05}. As usual, 
there are supporters for each of the different alternatives. Our point 
of view is that neither of them  should be neglected, for the following reasons. 
Imagine that we just choose to apply a verification tool directly to the obtained 
system implementation, and we indeed find security flaws. Now suppose that, while
trying to fix them, we realize that one (or several) of them is not just a coding
flaw, but a design flaw. That means that we must go back to the design phase and
fix the flaw at that stage. Moreover, if the flaw is serious enough, the existing
implementation may need extensive changes, which will incur in unacceptable costs
and delays. Yet another disadvantage is that code verification obviously relies
on the existence of a tool for verifying the specific programming language that
has been used. Although these tools are gaining popularity, the huge number of
existing programming languages, along with the complexity of creating such a tool,
suggests that, in the real world, we should not trust in having always a code
verification tool suitable to our needs. On the other hand, deciding just to apply 
a verification tool to test if our design is robust does not guarantee that the 
implementation of that design will also be secure even though the design seems to 
be so. However it is technology independent. Thus, we argue that verification of the 
security properties of both the design and its implementation should be applied 
when possible.

We propose here a simple, but yet powerful, methodology for the 
verification of the design of communication protocols. Thus, we keep in 
mind the specific risks that we can encounter when 
desiging them, but set aside the more general software flaws that, although 
can ultimately affect the security properties of the protocols, should be dealt with 
by the tools applied for code verification. As a methodology, its main advantage
is that it is directly devised to help detecting simple flaws early in the design
process, leaving only the complex, time-consuming ones for the last steps. Thus,
it avoids wasting too much time in flaws that could have been easily detected. It
also allows reaching the highest levels of assurance of other frameworks, like
\cite{mmob10,cc3}, depending on the tools used for procedural verification.
  
\section{The methodology}
\label{sec:methodology}

\begin{figure}[ht!]
\begin{center}
\includegraphics[width=0.85\textwidth]{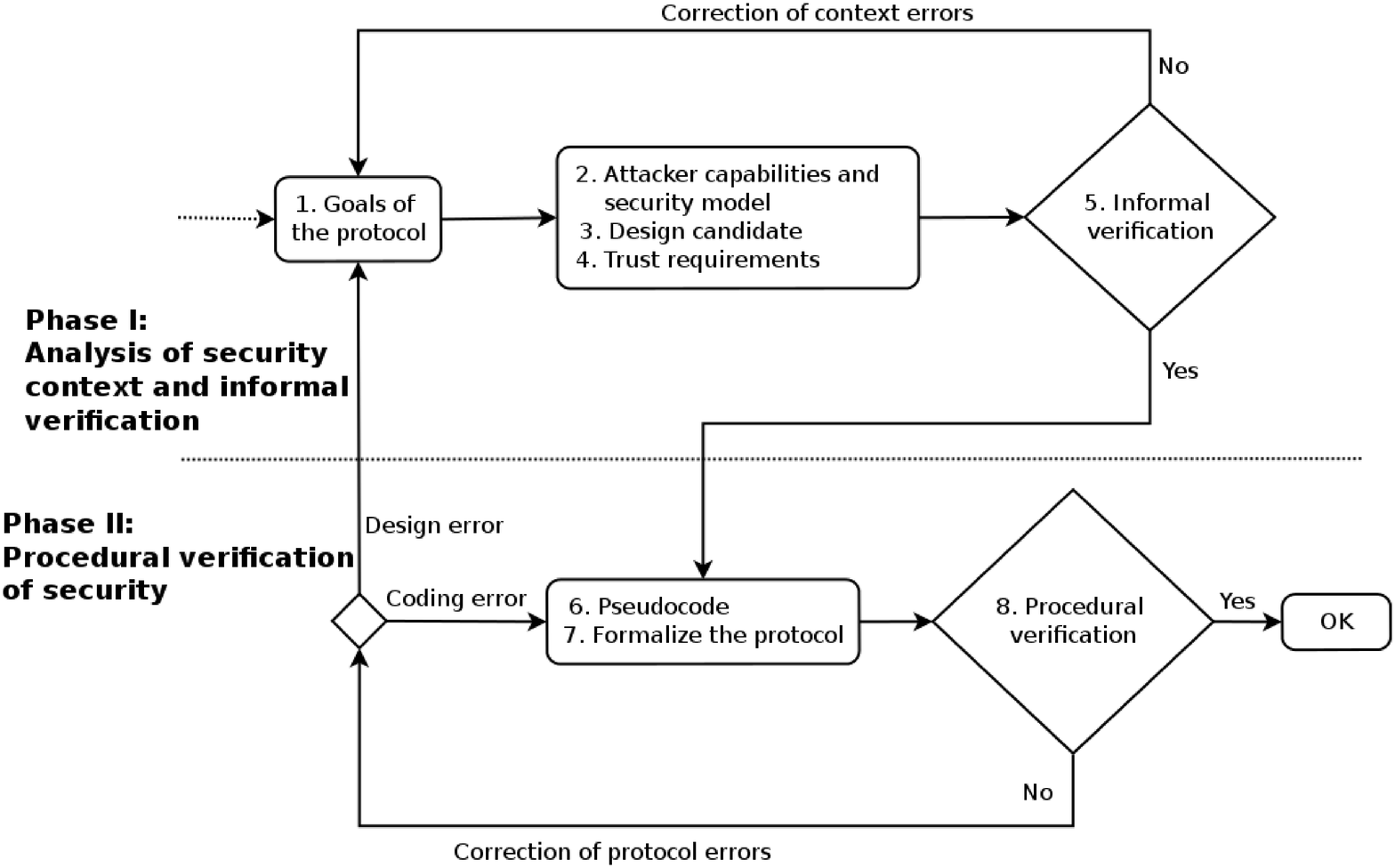}
\caption{Steps of the proposed methodology. From the initial design of
  the protocol, the first phase consists of performing an analysis of its
  goals, the security model that best fits our needs a design candidate, and 
  the trust requirements we expect to achieve. During the second phase, after giving
  informal and formal definitions of the protocol, we apply the chosen 
  tools or models for the verification of the security requirements 
  established before. Note that the methodology is structured 
  as a loop, which is repeated as long as new context, protocol errors 
  are detected.\label{fig:methodology}}
\end{center}
\end{figure}

Our methodology consists of two main parts, like shown in
Fig. \ref{fig:methodology}. The first part is concerned with specifying
the security requirements and considering if they are suitable in a given
context, and performing a first informal verification; while the second 
part is centered in  the formalization and formal verification of the protocol itself.

\subsection{Analysis of security context and informal verification}
\label{ssec:context}

This is the first part of the methodology and is mainly intended to avoid incongruences related 
to security requirements unachievable or inappropriate for some reason, and to 
improve our understanding of the protocol. It also allows us to perform several
informal checks of our first design candidates. This part is depicted in the 
first phase of Fig. \ref{fig:methodology}. Namely, here we face the following matters:

\begin{description}

\item[Step 1. Goals of the protocol.]

  This step might seem too obvious to be needed, but 
  specifying the goals of the protocol will help to improve the understanding 
  of the protocol, and will also come in handy for the third and
  fourth steps. As output here, we will get a rather
  informal specification of what we intend to achieve with our protocol,
  something like \emph{``Complete certainty that the person being registered
    in our website is who he/she says he/she is''}.
    
\item[Step 2. Attacker capabilities and security model.]

  This step is essential, because depending on the capabilities
  the attacker has, some security requirements might or might not be achievable.
  In this sense, there are two main attacker models: the Dolev-Yao attacker \cite{dy83}
  and the computational attacker \cite{bbm00}. The Dolev-Yao attacker model assumes
  perfect cryptographic primitives but gives otherwise full control to the attacker over 
  the communications network. The computational attacker is somehow more
  realistic in that it is assumed to be a Probabilistic Polynomial-time Turing Machine. 
  But, as a matter of fact, we are also worried about other attacker capabilities, like
  the capability to access public records containing personal information
  assumed to be secret in the protocol like Facebook pages, lists of personal 
  passwords published by hackers or leaked, etc.). Indeed, this kind
  of information is usually assumed to be secret in the typical
  attacker models, and it can drastically change the provided
  security guarantees. On the other hand, we might find 
  ourselves in a situation where all the capabilities of the typical Dolev-Yao 
  or computational attacker do not hold. So, in short, here we have to ask ourselves
  whether the assumptions made by standard attacker models are close to our 
  reality, or if we need to adjust them in some way. May we ignore this step,
  any later proof might be wrong because our initial assumptions were not close
  to our reality \cite{an95}. As a result, with this step we will get a specification of the attacker model,
  e.g. Dolev-Yao or computational, along with a list of specific capabilities added
  or subtracted to it, like \emph{``Knows the home address of everyone that may
    be involved in the protocol''} or \emph{``Cannot eavesdrop wired communications''}.
  Henceforth, we will refer to this list as the \emph{``+/- capabilities''} list.

\item[Step 3. Design candidate.] 

  Once we have made up our mind as to which are the goals of our protocol and the
  attacker capabilities and security model we assume, we can adventure ourselves
  to cast a design candidate. For communications protocols, a sequence diagram
  is most commonly used, since it captures both the order of the messages composing
  the protocol and their contents. Therefore, as output of this step we will 
  obtain a sequence diagram depicting the protocol.

\item[Step 4. Trust requirements.]
    
  This step of the methodology is the first part of an informal verification procedure. 
  Besides helping us in checking the security requirements, it is also intended to 
  prevent us from wasting our efforts needlessly. Before undertaking the formal 
  verification task, which may be time consuming, it is worth making a simpler 
  manual verification, in order to detect obvious mistakes. For that purpose, 
  we will follow the design candidate from the beginning, asigning trust properties. 
  Note that we use the term trust property rather than security property because
  at this point, this properties are something we expect the elements to keep, 
  not something we state that they have. The procedure is as follows: for each step 
  of the sequence diagram, we will note down the trust properties we expect from 
  every of its elements to keep. If, within a specific step, we apply some 
  function to merge several elements or disassemble them, we may have to require
  additional trust properties to the result (e.g., a message containing the
  a priori unauthenticated elements $x_1,x_2$ along with $MAC_K(x_1,x_2)$ can
  be considered authenticated, by the holder of a key $K$, if the MAC verification suceeds, provided that the
  key $K$ is trusted)\footnote{Typically, this kind of rules are specified formally
    inside formal verification tools (e.g., Cryptyc \cite{cryptyc} is based in
    type deduction rules). However, since this step is intended to be a preliminary 
    informal approach, common sense and experience are enough.}. Since the properties 
  we require from a specific element may change through the protocol, we will revise them 
  for every step, even if the element has already been processed previously. This process
  is summarized in Algorithm \ref{alg:requirements}. Also, the requirements that we may
  need to apply to the different elements are\footnote{Although we propose this 
    five requirements, more could be defined if necessary.}:

  \begin{description}
  \item[None.] When a field does not require any specific trust property.
  \item[Authenticity.] When the authenticity of the field must be guaranteed.
  \item[Confidentiality.] When the specific field must be kept secret.
  \item[Integrity.] When the integrity of the field must be preserved. Detecting
      any undesired modifications.
  \item[Uniqueness.] When an element cannot be repeated among the same or 
      different protocol runs. This property is commonly used to guarantee
      freshness and avoid replay attacks.
  \end{description}

  \begin{algorithm}[H]
    \SetAlgoLined   
    \KwData{The sequence diagram of the protocol.}
    \KwResult{A set of trust requirements for each protocol step.}
    \For{s = first step; until s = last step}{
      \For{e = first until e = last element of step s}{
        Set reqs[s][e];
      }
      reqs[s] += Additional trust requirements for step s;
    }
    \caption{Algorithm for assigning requirements to protocol elements and messages. With
      elements, we refer to any component calculated or sent within each specific step.
      \label{alg:requirements}}
  \end{algorithm}

  Therefore, as output of this step we get a list of trust requirements which,
  optionally, can be depicted jointly with the sequence diagram to create a requirement-tagged 
  sequence diagram of our protocol design candidate.

  Besides, we also have to decide in this step which 
  verification tool or process we will apply in the second phase. Now
  that we know our specific trust requirements, we can choose an appropriated tool, since
  some tools may not be applicable (depending on the security properties we want to verify).
  
\item[Step 5. Informal verification.]

  Now we have to take the +/- capabilities list and the 
  list of trust requirements (or the requirement tagged sequence diagram, whichever 
  representation we are using). If any of the trust requirements enters in conflict
  with any of the capabilities we granted to the attacker, then we have a security
  failure. If not, we have informally verified the design candidate and we can
  continue. However, we must keep in mind that this does not guarantee that the
  design will also pass the next phase of the methodology.
  
\end{description}

After applying our methodology up to this point for verifying a protocol, 
and assuming we passed the step 5, we would have reached the assurance level 
PAL1 defined in \cite{mmob10}.

\subsection{Procedural verification of security}
\label{ssec:procedural}

This part of the methodology is devoted to the procedural verification of the security
requirements established before. It is depicted in the second phase of Fig.
\ref{fig:methodology}. With procedural verification we refer to the fact that 
widely approved theories, methods or procedures should be applied here. Again, we can apply either the
formal or the computational model. The main advantage of the formal methods is that 
there are plenty of
tools that automatically analyze a protocol formalization. On the other hand, the
computational model verification takes into account that the cryptographic primitives
may not be perfect. However, in the last years, there has been a lot of work 
to prove that formal security implies computational security \cite{ar00,jlm04}, given
that the cryptographic primitives meet some properties. We do not enter
in this matter here, and just assume that a ``procedural'' verification model has
been adopted and is going to be applied. Indeed, the aim of our methodology
is to formally tackle the verification of the security properties of communication
protocols. Nevertheless, we have found that the application of automatic tools 
(like ProVerif \cite{blanchet10}, Cryptyc \cite{cryptyc}, etc.) allows to 
detect many flaws with little effort, so it may be a good choice to first apply 
formal methods and, if required, after that, use the computational model for 
more concrete evaluation.

Therefore, the steps we have included in this phase are as follows:

\begin{description}
  \item[\textbf{Step 6.}] \textbf{Protocol pseudocode}
    In the same way that writing pseudocode is useful before coding a program,
    writing an informal narration of a protocol in the shape of pseudocode, helps 
    to reach a higher concretion level before properly formalizing it. 
    As output of this step we get
    a written representation of the sequence diagram, with one process
    for each principal, which depicts the internal computations performed by
    each of them in order to generate the messages' components, along with the 
    messages sent to the other principals.
      
  \item[\textbf{Step 7.}] \textbf{Formalization of the protocol}
    From the protocol pseudocode, it is typically easy to produce a formalization
    in the language or definition model required by the tools we are going to use
    to verify the protocol (the ones we decided to use in step 4). The result of 
    this step must be a meticulous representation of how the protocol will be, once
    implemented.

  \item[\textbf{Step 8.}] \textbf{Procedural verification}
    The formalization obtained in the previous step will then be used as input
    for the chosen procedural verification model or tool. If the tools or procedures
    followed in this step ``output'' that all requirements are fulfilled, we can conclude. If not, we have to 
    go back to the first step and correct protocol errors in our design, checking also for coding
    errors in the formalization.

\end{description}

As an end note, we must point out that the methodology does not give
as output an explicit measure of security of the analysed protocol, other
  than checking whether the specified security requirements are held. That will depend
on the tools or the models used to verify the protocol. In any case, at least, we
will obtain an answer to whether or not the protocol meets the requirements specified
in step 4 of our methodology. Again referencing \cite{mmob10}, any 
protocol successfully verified using our methodology would reach an assurance 
level equivalent to PAL2 or PAL3, depending on whether the used tool provides
bounded or unbounded verification (or even the - yet under consideration in 
\cite{mmob10} - PAL4 level, would we have used a computational model verification 
tool).

To summarize, the inputs and outputs of each of the different steps of
our methodology are shown in Tables \ref{tab:iophase1} and \ref{tab:iophase2},
corresponding to the first and second phases of the methodology. The acronyms
used in the tables for the different outputs are explained in Table \ref{tab:outputs}.

\begin{table}[ht!]
  \begin{center}
    \begin{tabular}{r l}
      \hline
      O1:&Informal list of goals.\\
      O2:&Security model and +/- capabilities list.\\
      O3:&Sequence diagram.\\
      O4:&List of trust requirements / Requirement-tagged sequence diagram.\\
      O5:&Pseudocode.\\
      O6:&Formalization.\\
      \hline
    \end{tabular}
    \caption{Definition of the different outputs of the methodology.\label{tab:outputs}}
  \end{center}
\end{table}

\begin{table}[ht!]
  \begin{center}
    \begin{tabular}{|c|c|c|c|c|c|}
      \hline
      & \multirow{2}{*}{Goals} & Attacker model & Design & Trust & Informal\\
      & & and capabilities & candidate & requirements & verification\\
      \hline
      \textbf{Input:} & None & None & O1 & O3 & O2 and O4 \\
      \hline
      \textbf{Output:} & O1 & O2 & O3 & O4 & Yes/No\\
      \hline
    \end{tabular}
    \caption{Inputs and outputs of each step of the first phase of the methodology.\label{tab:iophase1}}
  \end{center}
\end{table}

\begin{table}[ht!]
  \begin{center}
    \begin{tabular}{|c|c|c|c|}
      \hline
      & \multirow{2}{*}{Pseudocode} & \multirow{2}{*}{Formalization} & Formal\\
      & & & verification\\
      \hline
      \textbf{Input:} & O4 & O2 and O5 & O6 \\
      \hline
      \textbf{Output:} & O5 & O6 & Yes/No\\
      \hline
    \end{tabular}
    \caption{Inputs and outputs of each step of the second phase of the methodology.\label{tab:iophase2}}
  \end{center}
\end{table}

\section{Case studies of real security protocols}
\label{sec:cases}

In this section we provide three case studies by applying our methodology
to real security protocols. For the first case, a problem is located in the 
first phase of the methodology, while in the second case, a problem is detected 
with a formal analysis in the second
phase.\footnote{It is not our purpose to present new flaws here. We rather intend 
to show how to avoid security design flaws using our methodology.} In the
other hand, in the third case, we go through the whole methodology to verify one 
more protocol \cite{dar11,dar11b} which successfully passes our tests.

\subsection{Case study I: context verification failure}
\label{ssec:casei}

\begin{figure}[ht!]
\begin{center}
\includegraphics[width=0.85\textwidth]{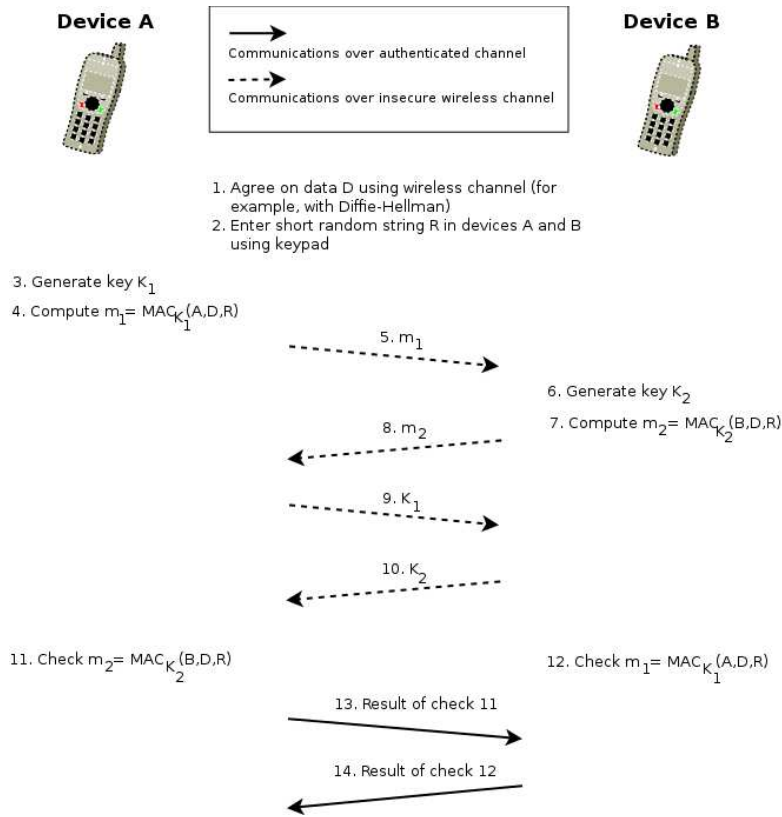}
\caption{Message sequence of the MANA III protocol \cite{gmn04}. 
  (1) Devices A and B agree on some data D (e.g., using Diffie-Hellman).
    (2) A secretly shared short random bitstring known in both devices with a keypad.
    (3-8) Both devices independently generate a key, compute a MAC of their identities,
    D and R and send the result to each other.
    (9-10) Both devices exchange the keys used in the previous MACs.
    (11-14) Both devices check if the received MACs are correct, and inform
    the other device of the obtained result.
    Messages 5,8,9 and 10 are sent over insecure channels, while messages
    13 and 14 are sent over an authenticated channel.\label{fig:mana-sequence}}
\end{center}
\end{figure}

Here we apply the first part of the methodology to the MANA III (MANual Authentication) 
authentication protocol \cite{gmn04}. MANA III is intended for wireless networks and
bases its security in the requirement of manually introducing a short bitstring (R in Fig. 
\ref{fig:mana-sequence}) via keypads, previous to the execution, which is kept secret. 
Since this short secret bitstring is used in the calculation of MACs, an attacker will 
not be able to create fake MACs for authentication in real time. The MANA III protocol is 
informally depicted in Fig. \ref{fig:mana-sequence}.

In \cite{ws05} the authors suggest that this authentication method is no longer
secure due to the proliferation of CCTV cameras. Indeed, such cameras can be
used to observe the short bitstring introduced via the keypad. If the secret short 
random bitstring becomes known to the attacker, he could use it to mount a Man In 
The Middle attack \cite{ws05}. Let us now see how this fact could have been detected with
the first part of our methodology.

\begin{description}
\item[Step 1. Goals of the protocol.]

  We want to achieve a final state in wich Device A and Device B are 
  certain that they are communicating with each other. That is directly
    translated to authenticity. Thus, our specific goals are:

  \begin{enumerate}
  \item Device A is successfully authenticated against Device B.
  \item Device B is successfully authenticated against Device A.
  \end{enumerate}

\item[Step 2. Attacker capabilities and security model.]

  Here we have to decide an attacker model, let us assume that we adopt the Dolev-Yao attacker 
  model (nevertheless, this decision does not affect what follows). As we said 
  in Sec. \ref{ssec:context} we have to analyze our specific context to see if our 
  attacker would have some extra capabilities (or if we have to take out some capability
  from him). In the MANA III protocol, and given the current state of development of
  CCTV cameras, like it is said in \cite{ws05}, it is not safe to assume that, in
  any situation, a password introduced via a keypad will always remain secret. 
  One could also make a quick search on the Internet and find 
  several microcameras of roughly 15x5 milimiters. With cameras that small, an
  attacker could manage to record almost anything without the victim even noticing.
  Therefore, the more realistic decision of giving the attacker the capability to 
  observe keyed passwords seems to be justified\footnote{However, this may be a too
    restrictive scenario for many situations, but for illustration purposes, we will
    assume that we intend to use MANA-III in a security critical context.}. Besides, we will
  trust that the assumption made by the authors of the protocol stating that
  the short random string R prevents attackers from finding, in real time, 
  collisions for the calculated MACs. Summarizing, we use the Dolev-Yao
  model, and our +/- capability list will be as shown in Listing \ref{lst:+-mana}
  
  \begin{lstlisting}[frame=single,captionpos=b,
    caption={+/- capabilities list for the Dolev-Yao attacker for the MANA III
      protocol.},
    label={lst:+-mana},
    escapechar=\%]
    % $+$% %May observe keystrokes.%
    % $-$% %Cannot find MAC collisions in real-time without R.%
  \end{lstlisting}
  
\item[Step 3. Design candidate.] From the goals we defined in the first step, we
  come up with a design candidate in the shape of a sequence diagram. Suppose that
  we produce the diagram in Fig. \ref{fig:mana-sequence}\footnote{Obviously, in a 
    real scenario, we will produce it at this step, and not before. We just showed
    the diagram in advance for introducing the case study.}.
  
\item[Step 4. Trust requirements.]

  With the sequence
  diagram in front of us, we apply Algorithm \ref{alg:requirements}. As a result,
  we obtain the list of trust requirements shown in Listing \ref{lst:reqcase1} (we omit 
  the complete process for brevity).

  \begin{lstlisting}[frame=single,captionpos=b,
    caption={List of trust requirements for MANA III.},
    label={lst:reqcase1},
    escapechar=\%]
        %\textbf{Initialization}%. A and B are publicly known values.
        %\textbf{Step 1}%. D : Authenticity
        %\textbf{Step 2}%. R : Confidentiality
        %\textbf{Step 3}%. %$K_1$% : Confidentiality
        %\textbf{Step 4}%. %$m_1 = MAC_{K_1}(A,D,R)$% : Authenticity 
                ( %$K_1$% %is% trusted by A since she has 
                  created it. )
        %\textbf{Step 5}%. %$m_1$% : None 
                ( At this point, no one knows %$K_1$% and %$m_1$% 
                  %is% sent over the %public% channel, so we 
                  cannot place any trust %in% this element. )
        %\textbf{Step 6}%. %$K_2$% : Confidentiality                                            
        %\textbf{Step 7}%. %$m_2 = MAC_{K_2}(B,D,R)$% : Authenticity 
                ( %$K_2$% %is% trusted by B since she has 
                  created it. )
        %\textbf{Step 8}%. %$m_1$% : None 
                ( At this point, no one knows %$K_2$% and %$m_2$% 
                  %is% sent over the %public% channel, so we 
                  cannot place any trust %in% this element. )
        %\textbf{Step 9}%. %$K_1$% : None 
                ( The used cannel %is% insecure, anyone can 
                  %send% this. )
        %\textbf{Step 10}%. %$K_2$% : None 
                ( The used cannel %is% insecure, anyone can 
                  %send% this. )
        %\textbf{Step 11}%. Nothing to do here.
        %\textbf{Step 12}%. Nothing to do here.
        %\textbf{Step 13}%. %$check_{m_2}$% : Authenticity.
        %\textbf{Step 14}%. %$check_{m_1}$% : Authenticity.
  \end{lstlisting}

\item[Step 5. Informal verification.]

  Going through the list of trust requirements shown in Listing \ref{lst:reqcase1}
  we soon find an incompatibility with the attacker capabilities: in the second step,
  we see that we require R to be secret (confidentiality). Still, we gave the attacker
  the capability to observe keystrokes, and since R is entered in Devices A and B
  via a keypad, we cannot consider it secret.
  
\end{description}

We should see now the importance of this part of the methodology, and why does
it have to be applied with care. Although the standard attacker models
provide a powerful framework to start with, there are cases where these models
do not cover all the possibilities that the attacker may have available. Therefore, in order
to detect possible flaws on the context of the protocol, we have to carefully
establish the protocol goals, analyze the attacker capabilities given the current 
technologies, give a formal statement of the required security properties, and check 
if they hold given the previous facts. Once we successfully complete
this phase, we can proceed with the second phase of the methodology, with a higher
degree of certainty that we are correctly grounded the context of our protocol.

\subsection{Case study II: procedural security verification failure}
\label{ssec:caseii}

We dedicate this subsection to the procedural analysis of the Shared Key
Authentication of the WEP standard (WEP-SKA from now on), as defined in \cite{IEEE802.11}. 
Let us then assume that the protocol successfully passes the first
phase of our methodology. In this
case, we apply the automatic formal verifier ProVerif \cite{blanchet10}. 

\begin{figure}[ht!]
\begin{center}
\includegraphics[width=0.75\textwidth]{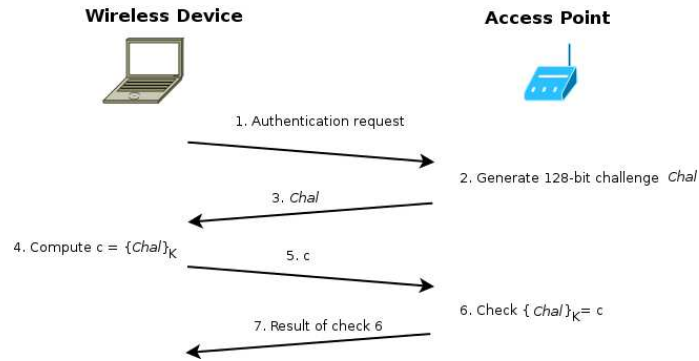}
\caption{Message sequence of the WEP-SKA protocol \cite{IEEE802.11}. 
  (1) The WD requests to be authenticated using WEP-SKA.
  (2-3) The AP generates a challenge, and sends it, in plaintext, to WD.
  (4-5) The WD encrypts the challenge with a preshared key, and sends
        the result to the AP.
  (6-7) The checks the received encrypted challenge and informs WD of
        the result.\label{fig:wep-ska-sequence}}
\end{center}
\end{figure}

The WEP-SKA protocol is one of the two authentication methods supported by the
WEP standard. A normal execution consists of four messages, between two
stations. We will call them the Wireless Device (WD), which wants to be authenticated,
and the Access Point (AP) which attends the WD's request. A typical protocol run
is depicted in Fig. \ref{fig:wep-ska-sequence}.

Nevertheless, as pointed out in the WEP standard \cite{IEEE802.11} sending both the
challenge and its encrypted version may produce a security problem. This is
due to the fact that the encryption algorithm used (RC4) is 
a stream cipher which generates a pseudorandom sequence and XORs it to the
plaintext in order to create the ciphertext. Therefore, if we know the 
ciphertext and the plaintext, we can obtain the
keystream straightaway. Nevertheless, in the standard
it was only advised (but not required) to change the key and/or IV 
(Initialization Vector) frequently. Therefore,
as observed in \cite{bgw01} the fact of not being required to change the key/IV,
indirectly forces every receiver to accept repeated key/IV's, or risk otherwise
not being compatible with some WEP compliant devices. That allows an attacker to 
successfully impersonate any station after having observed one single authentication. 
In  \cite{bgw01} they call this attack \emph{Authentication spoofing}.

Let us now forget for a moment that this flaw is known, suppose that we
are asked to verify the security of WEP-SKA, and that it successfully
passes the first part of our methodology. Then we would have to face the procedural 
verification of the protocol. The steps defined in Sec. \ref{ssec:procedural}
are as follows:

\begin{description}

\item[Step 6. Protocol pseudocode.]

    The pseudocode for the WEP-SKA protocol is given in Listing 
    \ref{lst:wepska-informal}.


    \begin{lstlisting}[frame=single,captionpos=b,
      caption={Pseudocode for the WEP-SKA procedures. $WD$ stands from the
        Wireless Device to be authenticated, $AP$ stands from Acces Point
        and $k$ is the shared key. For simplicity, it is only shown 
        the pseudocode for a successful execution. The field \emph{info} conveys
        information dependent on the authentication algorithm, and \emph{result} stores 
        the result of the requested authentication. \emph{MA, AUTH, 'shared key'} and the
        numbers used in the messages are fields specified in the standard, see 
        \cite{IEEE802.11}},
      label={lst:wepska-informal},
      escapechar=\%]
      Process %\textbf{WD}%:
         send(MA,AUTH,(id,'shared key',1));
         receive(MA,AUTH,('shared key',2,info,result));
         if result == sucessful then
             chal = info;
             send(MA,AUTH,('shared key',3,%$\{chal\}_k$%));
         fi
         receive(MA,AUTH,('shared key',4,result));

      Process %\textbf{AP}%:
         receive(MA,AUTH,(id,'shared key',1));
         if successful then
             result = sucessful;
             chal = new pseudorandom;
             send(MA,AUTH,('shared key',2,chal,result));
             receive(MA,AUTH,('shared key',3,%$\{chal2\}_k$%));
             if %$\{chal2\}_k$% == %$\{chal\}_k$% then
                 send(MA,AUTH,('shared key',4,sucessful))
             fi
         fi        
    \end{lstlisting}
    
    
  \item[Step 7. Formalization of the protocol.]

    ProVerif has a problem here, 
    because it does not support the XOR opperation\footnote{Although there are
      approaches for XOR-aware modifications of ProVerif. See \cite{kt08}.}.
    Nevertheless, since we don't need to apply complex derivation rules,
    we can get around this problem with a simple reduction rule simulating the
    XOR. This workaround can be seen in the code of the program available in \cite{code:wepska},
    and is a rule telling that if the attacker knows both
    the challenge $c$, and the challenge $c$ encrypted (i.e., xored) with the key 
    $k$, then he can apply the xor function to recover the key $k$. 

  \item[Step 8. Procedural verification.]

    Once we have formalized the protocol, we run the tool to see if it 
    succeeds or fails. In our case, running ProVerif with the code in 
    \cite{code:wepska}, we observe a trace like the one shown in Listing 
    \ref{lst:wepskatrace}.
    

    \begin{lstlisting}[frame=single,captionpos=b,
      caption={An example attack trace found by ProVerif for the Authentication spoofing
        attack over WEP-SKA introduced in \cite{bgw01}. In the first block,
        the attacker eavesdrops on a WEP-SKA run between a legitimate WD and the AP. As
        a result, the attacker obtains the preshared key $k$. In the second block, the
        attacker uses the key $k$ in order to successfully complete a WEP-SKA execution
        with the AP.},
      label={lst:wepskatrace},
      escapechar=\%]
        % \textbf{WEP-SKA between legitimate $WD_1$ and $AP$}%:
        % $WD_1 \rightarrow  AP $% : %$WD_1$%
        % $AP \rightarrow WD_1 $% : %$Chall_1$%
        % $WD_1 \rightarrow AP $% : %$\{Chall_1\}_k$%
        Attacker : %$k = Chall_1 \oplus \{Chall_1\}_k$% %$\implies$% Attacker gains %$k$%
        % $AP \rightarrow WD_1 $% : OK
        % \textbf{WEP-SKA between illegitimate $fake_{WD_2}$ and AP}%:
        % $fake_{WD_2} \rightarrow AP $%: %$fake_{WD_2}$%
        % $AP \rightarrow fake_{WD_2}$% : %$Chall_2$%
        % $fake_{WD_2} \rightarrow AP $%: %$\{Chall_2\}_k$%
        % $AP \rightarrow fake_{WD_2}$% : OK
    \end{lstlisting}
    
    Since we have found an attack during the procedural verification with ProVerif, 
    the requirements are not fulfilled. Therefore, we must go back to the first stage of the methodology 
    once we have checked that no \emph{coding} error has been made and redesign the protocol 
    to avoid this attack. For instance, we could proceed like the authors
    of \cite{bgw01} suggest, which is to disallow the reuse of IVs in order to 
    avoid this attack.
    
  \end{description}
  
\subsection{Case study III: complete verification}
\label{ssec:caseiii}

Now we shortly show how the full methodology is applied to the analysis of
the protocol that was informally presented in 
\cite{dar11}, and whose formal security verification is performed in \cite{dar11b}. While
applying our methodology, we found flaws in the design of the protocol that could
lead to, for instance, replay attacks. Here we will just summarize
the procedure followed (for further details, see \cite{dar11b}).
The protocol is called CHAT-SRP (CHAos based Tickets - Secure Registration Protocol), 
and it is intended to be used in registration protocols for interactive platforms. 
The typical registration method used in these platforms requires the new users to 
provide an email address, verifying their identity when they
access a link included in an email sent to the provided address\footnote{This
is known as EBIA, \cite{garfinkel03}.}. This approach, although very user-friendly,
is very insecure, because emails are almost always sent unprotected \cite{farrell11}. 
Therefore, an attacker may impersonate a user just by
eavesdropping the activation email. CHAT-SRP generates a ticket (basically, a pseudo-random
number), links it to the requesting user email, and sends it encrypted
via HTTPS. The user, who is initially identified as being the owner of a mobile number
(which we assume to be previously known and verified),
besides accessing the activation link sent by email, has to
provide the right ticket in order to validate his new account. This way we 
confirm that who requested the registration is the one completing it. Once
confirmed his identity, the protocol sends the new user a digital identity, which 
he could later use for performing robust cryptographic operations. 
Let us use our methodology to analyze it.


\begin{description}
\item[Step 1. Goals of the protocol.]

  We want to register new users, providing them with new 
  virtual identities to be used in the platform. Therefore, our 
  more formally stated goals are:

  \begin{enumerate}
  \item Authenticate a user requesting registration.
  \item Provide him/her a digital identity, keeping the identity's confidentiality
    and guaranteeing its authenticity (i.e., that it has been generated
    by a suitable Certification Authority).
  \end{enumerate}

\item[Step 2. Attacker capabilities and security model.]

  We assume the typical of the Dolev-Yao attacker. Note that 
  this includes the capability of eavesdropping and blocking emails, which, 
  as we said before, is our main concern. Besides, we also assume that our
  attacker can search in the Internet for the necessary information to start
  a registration process (in this case, a username and an email). Thus, our
  +/- capabilities list is shown in Listing \ref{lst:+-chatsrp}.

  \begin{lstlisting}[frame=single,captionpos=b,
    caption={+/- capabilities list for the Dolev-Yao attacker for the CHAT-SRP
      protocol.},
    label={lst:+-chatsrp},
    escapechar=\%]
    % $+$% %Knows all the information required for registration
          for every user subject to be registered (username and email).%
  \end{lstlisting}

\item[Step 3. Design candidate.]

  A sequence diagram for the protocol informally described at the 
  beginning of this subsection is shown in Fig. \ref{fig:sequence},
  and we will use it as our design candidate. Keep in mind that, for brevity, we only show the result of
  several iterations of the methodology. Also, in this diagram and in the
  subsequent analysis, WS stands for Web Server, RA for Registration Authority
  and CA for Certification Authority. The three of them are assumed to be
  trusted authorities.
  
  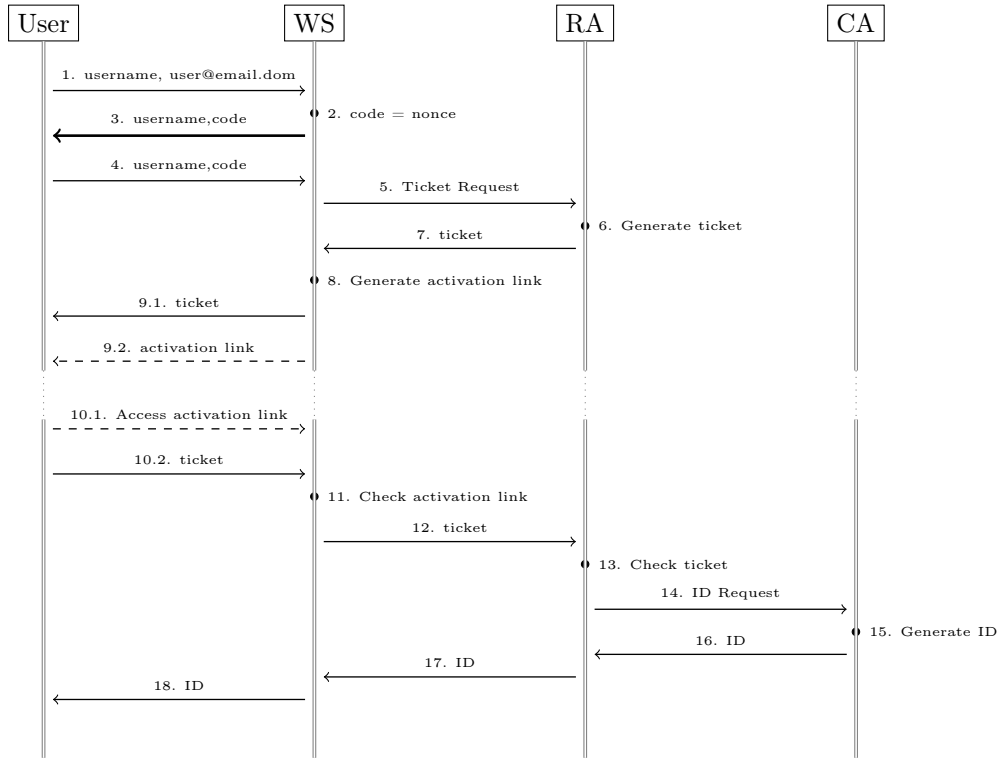
\begin{figure}[h!]
    \begin{center}

\begin{tikzpicture}[scale=1.2]
  
  \node (user) at (1,8.25) [draw] {User};
  \node (user1) at (1,7.50) [draw=none] {};
  \node (user2) at (1,7.00) [draw=none] {};
  \node (user3) at (1,6.50) [draw=none] {};
  \node (user4) at (1,6.25) [draw=none] {};
  \node (user5) at (1,5.75) [draw=none] {};
  \node (user6) at (1,5.25) [draw=none] {};
  \node (user7) at (1,5.00) [draw=none] {};
  \node (user8) at (1,4.5) [draw=none] {};
  \node (user9) at (1,3.75) [draw=none] {};
  \node (user10) at (1,3.25) [draw=none] {};
  \node (user11) at (1,2.50) [draw=none] {};
  \node (user12) at (1,1.75) [draw=none] {};
  \node (user13) at (1,1.25) [draw=none] {};
  \node (user14) at (1,1.00) [draw=none] {};
  \node (user15) at (1,0.75) [draw=none] {};
  \node (enduser) at (1,0.00) {};
  
  \node (ws) at (4,8.25) [draw] {WS};
  \node (ws1) at (4,7.50) [draw=none] {};
  \node[circle, scale=0.3, fill=black] (wsgencodeline) at (4,7.25) [draw] {};
  \node (wsgencode) [right=0 of wsgencodeline] { \tiny{2. code = nonce} };
  \node (ws2) at (4,7.00) [draw=none] {};
  \node (ws3) at (4,6.50) [draw=none] {};
  \node (ws4) at (4,6.25) [draw=none] {};
  \node (ws5) at (4,5.75) [draw=none] {};
  \node[circle, scale=0.3, fill=black] (wsgenlinkline) at (4,5.40) [draw] {};
  \node (wsgenlink) [right=0 of wsgenlinkline] { \tiny{8. Generate activation link} };
  \node (ws7) at (4,5.00) [draw=none] {};
  \node (ws8) at (4,4.5) [draw=none] {};
  \node (ws9) at (4,3.75) [draw=none] {};
  \node (ws10) at (4,3.25) [draw=none] {};
  \node[circle, scale=0.3, fill=black] (wschecklinkline) at (4,3) [draw] {};
  \node (wschecklink) [right=0 of wschecklinkline] { \tiny{11. Check activation link} };
  \node (ws11) at (4,2.50) [draw=none] {};
  \node (ws12) at (4,1.75) [draw=none] {};
  \node (ws13) at (4,1.25) [draw=none] {};
  \node (ws14) at (4,1.00) [draw=none] {};
  \node (ws15) at (4,0.75) [draw=none] {};
  \node (endws) at (4,0.00) {};

  \node (ra) at (7,8.25) [draw] {RA};
  \node (ra1) at (7,7.50) [draw=none] {};
  \node (ra2) at (7,7.00) [draw=none] {};
  \node (ra3) at (7,6.50) [draw=none] {};
  \node (ra4) at (7,6.25) [draw=none] {};
  \node[circle, scale=0.3, fill=black] (ragenticketline) at (7,6.00) [draw] {};
  \node (ragenticket) [right=0 of ragenticketline] { \tiny{6. Generate ticket} }; 
  \node (ra5) at (7,5.75) [draw=none] {};
  \node (ra6) at (7,5.25) [draw=none] {};
  \node (ra7) at (7,5.00) [draw=none] {};
  \node (ra8) at (7,4.5) [draw=none] {};
  \node (ra9) at (7,3.75) [draw=none] {};
  \node (ra10) at (7,3.25) [draw=none] {};
  \node (ra11) at (7,2.50) [draw=none] {};
  \node[circle, scale=0.3, fill=black] (racheckticketline) at (7,2.25) [draw] {};
  \node (rackeckticket) [right=0 of racheckticketline] { \tiny{13. Check ticket} };
  \node (ra12) at (7,1.75) [draw=none] {};
  \node (ra13) at (7,1.25) [draw=none] {};
  \node (ra14) at (7,1.00) [draw=none] {};
  \node (ra15) at (7,0.75) [draw=none] {};
  \node (endra) at (7,0.00) {};

  \node (ca) at (10,8.25) [draw] {CA};
  \node (ca1) at (10,7.50) [draw=none] {};
  \node (ca2) at (10,7.00) [draw=none] {};
  \node (ca3) at (10,6.50) [draw=none] {};
  \node (ca4) at (10,6.25) [draw=none] {};
  \node (ca5) at (10,5.75) [draw=none] {};
  \node (ca6) at (10,5.25) [draw=none] {};
  \node (ca7) at (10,5.00) [draw=none] {};
  \node (ca8) at (10,4.5) [draw=none] {};
  \node (ca9) at (10,3.75) [draw=none] {};
  \node (ca10) at (10,3.25) [draw=none] {};
  \node (ca11) at (10,2.50) [draw=none] {};
  \node (ca12) at (10,1.75) [draw=none] {};
  \node[circle, scale=0.3, fill=black] (cagenidline) at (10,1.50) [draw] {};
  \node (cagenid) [right=0 of cagenidline] { \tiny{15. Generate ID} };
  \node (ca13) at (10,1.25) [draw=none] {};
  \node (ca14) at (10,1.00) [draw=none] {};
  \node (ca15) at (10,0.75) [draw=none] {};
  \node (endca) at (10,0.00) {};

  \draw[-] (user.south) -- (user8.south) [double,color=gray];
  \draw[-] (user8.south) -- (user9.north) [dotted,color=gray];
  \draw[-] (user9.north) -- (enduser.north) [double,color=gray];
  \draw[-] (ws.south) -- (ws8.south) [double,color=gray];
  \draw[-] (ws8.south) -- (ws9.north) [dotted,color=gray];
  \draw[-] (ws9.north) -- (endws.north) [double,color=gray];
  \draw[-] (ra.south) -- (ra8.south) [double,color=gray];
  \draw[-] (ra8.south) -- (ra9.north) [dotted,color=gray];
  \draw[-] (ra9.north) -- (endra.north) [double,color=gray];
  \draw[-] (ca.south) -- (ca8.south) [double,color=gray];
  \draw[-] (ca8.south) -- (ca9.north) [dotted,color=gray];
  \draw[-] (ca9.north) -- (endca.north) [double,color=gray];

  \draw[->] (user1) -- (ws1) [line width=0.5pt] node [midway,above,draw=none]{\tiny{1. username, user@email.dom}};
  \draw[->] (ws2) -- (user2) [line width=1pt] node [midway,above,draw=none]{\tiny{3. username,code}};
  \draw[->] (user3) -- (ws3) [line width=0.5pt] node [midway,above,draw=none]{\tiny{4. username,code}};
  \draw[->] (ws4) -- (ra4) [line width=0.5pt] node [midway,above,draw=none]{\tiny{5. Ticket Request}};
  \draw[->] (ra5) -- (ws5) [line width=0.5pt] node [midway,above,draw=none]{\tiny{7. ticket}};
  \draw[->] (ws7) -- (user7) [line width=0.5pt] node [midway,above,draw=none]{\tiny{9.1. ticket}};
  \draw[->] (ws8) -- (user8) [line width=0.5pt,dashed] node [midway,above,draw=none]{\tiny{9.2. activation link}};
  \draw[->] (user9) -- (ws9) [line width=0.5pt,dashed] node [midway,above,draw=none]{\tiny{10.1. Access activation link}};
  \draw[->] (user10) -- (ws10) [line width=0.5pt] node [midway,above,draw=none]{\tiny{10.2. ticket}};
  \draw[->] (ws11) -- (ra11) [line width=0.5pt] node [midway,above,draw=none]{\tiny{12. ticket}};
  \draw[->] (ra12) -- (ca12) [line width=0.5pt] node [midway,above,draw=none]{\tiny{14. ID Request}};
  \draw[->] (ca13) -- (ra13) [line width=0.5pt] node [midway,above,draw=none]{\tiny{16. ID}};
  \draw[->] (ra14) -- (ws14) [line width=0.5pt] node [midway,above,draw=none]{\tiny{17. ID}};
  \draw[->] (ws15) -- (user15) [line width=0.5pt] node [midway,above,draw=none]{\tiny{18. ID}};


  
\end{tikzpicture}
      \caption{Sequence diagram of the protocol messages. The dashed
        lines represent unprotected communications; the thin continuous ones represent SSL
        protected communications, and the thick continuous line represents the message sent
        using the extra authenticated channel (SMS).  Although not shown in the diagram
        for readability, the ticket is generated as the hash of a nonce and the user's email.
        \label{fig:sequence}}
    \end{center}
  \end{figure}  
  
\item[Step 4. Trust requirements.]
  
  Applying Algorithm \ref{alg:requirements} to the elements in diagram in Fig. 
  \ref{fig:sequence}, we obtain the trust requirements in Listing \ref{lst:reqcase3}.

  \begin{lstlisting}[frame=single,captionpos=b,
    caption={List of trust requirements for CHAT-SRP.},
    label={lst:reqcase3},
    escapechar=\%]
        %\textbf{Step 1}%. username : None 
               user@email.dom : None
        %\textbf{Step 2}%. code : Uniqueness, authenticity, confidentiality 
        %\textbf{Step 3}%. username : None
               code : Uniqueness, authenticity, confidentiality
        %\textbf{Step 4}%. username : None
               code : Uniqueness, authenticity, confidentiality
        %\textbf{Step 5}%. Ticket Request : Authenticity, confidentiality
        %\textbf{Step 6}%. ticket : Authenticity, confidentiality, uniqueness
        %\textbf{Step 7}%. ticket : Authenticity, confidentiality, uniqueness
        %\textbf{Step 8}%. link : Authenticity, confidentiality, uniqueness
        %\textbf{Step 9.1}%. ticket : Authenticity, confidentiality, uniqueness
        %\textbf{Step 9.2}%. link : Authenticity, uniqueness
        %\textbf{Step 10.1}%. link : None
        %\textbf{Step 10.2}%. ticket : Authenticity, confidentiality, uniqueness
        %\textbf{Step 11}%. Nothing to do here
        %\textbf{Step 12}%. ticket : Authenticity, confidentiality, uniqueness
        %\textbf{Step 13}%. Nothing to do here
        %\textbf{Step 14}%. ID Request : Authenticity, confidentiality
        %\textbf{Step 15}%. ID : Authenticity, confidentiality, uniqueness
        %\textbf{Step 16}%. ID : Authenticity, confidentiality, uniqueness
        %\textbf{Step 17}%. ID : Authenticity, confidentiality, uniqueness
        %\textbf{Step 18}%. ID : Authenticity, confidentiality, uniqueness
  \end{lstlisting}

  Basically, at steps 1, 3 and 4, we are considering the extra capability added
  to our attacker in our +/- capabilities list. Also, note that the activation 
  link, when created in step 8 is 
  required authenticity and uniqueness (the confidentiality requirement is just 
  a consequence of having just been created).
  However, in step 9.2 it is sent via email (an insecure channel), thus confidentiality
  cannot be assumed. Moreover, the link is sent at step 10.2 (which simulates the
  access to the corresponding web site) cannot be assumed to be confidential,
  nor unique, nor authenticated, since it can be originated from anywhere. The
  ticket and ID are expected to be kept authenticated, secret and unique, since
  they are always conveyed through SSL-secured channels.

  Since we have required the protocol to provide secrecy, authenticity 
  and uniqueness, we decided to use ProVerif, which allows to check all the 
  stated properties.

\item[Step 5. Informal verification.]

  Going through all the steps in Listing \ref{lst:reqcase3} we do not find
  any inconsistency with the attacker capabilities nor the +/- capabilities list
  specified before. Moreover, we have taken into account the attacker's capability
  to know all the personal information required to initiate a registration request
  (since we have place no trust requirement in the username and email elements in
  steps 1, 3 and 4). For the remaining steps, the required trust properties are
  consistent with the properties of the channels used to conveyed the associated
  elements (we must remark here that the WS, RA and CA are trusted authorities).
  Thus, we can continue, considering that our design successfully passes our
  informal verification step.

\item[Step 6. Protocol seudocode.]
  
  For brevity, we do not include 
  the pseudocode here. Instead, it is accessible from \cite{code:chatsrp}. Note 
  that, again, it was obtained after several iterations of our  methodology, 
  in which we detected and corrected security flaws.

\item[Step 7. Formalization of the protocol.]
  
  In step 4 we decided to use ProVerif for verification. From the 
  pseudocode, it is not hard to obtain a formalization for ProVerif. For further 
  details on the formalization of the protocol, see \cite{dar11b} or the pseudocode 
  and formalization for ProVerif in \cite{code:chatsrp}.

\item[Step 8. Procedural verification.]
  
  Running ProVerif on the code in \cite{code:chatsrp} shows no violation on the 
  security requirements. Therefore, we can assume that they have been accomplished.  

\end{description}

Therefore, following our methodology, as we have briefly shown, CHAT-SRP is seen to 
fulfill our secrecy, authenticity and uniqueness requirements. 

\section{Conclusion}
\label{sec:conclusion}

In this work we have designed a methodology for verifying security
properties of communication protocols. It is mainly divided in two parts, the
first one devoted to a preliminary study of the properties of the context where
the protocol will be applied, its security requirements and an informal
verification. The second part is dedicated to a procedural verification of these 
security requirements, applying well-known and widely accepted tools and/or 
procedures. By adopting an iterative methodology in which the early
processes are less time consuming, we avoid wasting resources by increasing the
probability of detecting basic flaws at the beginning of the analysis.
We also provide three examples in order to help to 
understand which are the advantages of applying this methodology. These examples 
(MANA III, WEP-SKA and CHAT-SRP) are representative, since they are real protocols 
that have been presented to the community, and specially in the case of WEP-SKA, widely used.
Note that the first two examples study already known security flaws.
However, our aim is not to detect new flaws, but to show how our methodology 
can be applied and the benefits of doing so.

We would like to remark that, when designing and evaluating communication protocols,
the application of such a methodology helps to detect and avoid flaws that
can lead to attacks on the protocols, like we have shown. It is important 
to note that both phases of the methodology are required: if the 
first phase is skipped, a wrong contextualization may render the formalization invalid; 
if, in the other hand, we skip the second phase, most probably we will not realize several involved 
nuances that, in turn, can lead to security flaws. By using what we called 
procedural methods, like formal protocol verifiers, or the computational model 
for the verification of protocols, we can prevent much of the damage caused by 
flawed designs. Nevertheless, the human factor also intervenes in the procedural
analysis of protocols, and a wrong formalization can make some flaws to pass 
unnoticed. Hence, like in every engineering process, this does not provide a 
$100\%$ success rate, but it does help to avoid many flaws.


\bibliographystyle{plain}
\bibliography{methodology}

\end{document}